\documentclass[prb,aps,twocolumn,showpacs,citeautoscript,longbibliography,superscriptaddress]{revtex4-1}

\usepackage{amsmath}
\usepackage{bm}
\usepackage{amssymb}
\usepackage{graphicx}
\usepackage[colorlinks=true,allcolors=blue]{hyperref}

\begin{document}

\title{Dynamical structure factor in the non-Abelian phase of the Kitaev honeycomb model in the presence of quenched disorder}

\author{Daniel Otten}
\affiliation{JARA-Institute for Quantum Information, RWTH Aachen University,
D-52056 Aachen, Germany}

\author{Ananda Roy}
\affiliation{JARA-Institute for Quantum Information, RWTH Aachen University,
D-52056 Aachen, Germany}
\affiliation{Institut de Physique Th\'eorique, Paris Saclay University, CEA, CNRS, F-91191 Gif-sur-Yvette.}
\author{Fabian Hassler}
\affiliation{JARA-Institute for Quantum Information, RWTH Aachen University,
D-52056 Aachen, Germany}

\begin{abstract} 
Kitaev's model of spins interacting on a honeycomb lattice describes a quantum spin-liquid, where an emergent static $\mathbb{Z}_2$ gauge field is coupled to Majorana fermions. In the presence of an external magnetic field and for a range of interaction strengths, the system behaves as a gapped, non-Abelian quantum spin-liquid. In this phase, the vortex excitations of the emergent $\mathbb{Z}_2$ gauge field have Majorana zero modes bound to them. Motivated by recent experimental progress in measuring and characterizing real materials that could exhibit spin-liquid behavior, we analytically calculate the dynamical spin structure factor in the non-Abelian phase of the Kitaev's honeycomb model. In particular, we treat the case of quenched disorder in the vortex configurations. Our calculations reveal a peak in the low-energy dynamical structure factor that is a signature of the spin-liquid behavior. We map the effective Hamiltonian to that of a chiral \textit{p}-wave superconductor by using the Jordan-Wigner transformation. Subsequently, we analytically calculate the wave functions of the Majorana zero modes, the energy splitting for finite separation of the vortices and finally, the dynamical structure factor in presence of quenched disorder.
\end{abstract}

\maketitle
\section{Introduction}

Quantum information processing using creation and manipulation of topological excitations in low-dimensional systems have been the focus of intense investigations lately. These topological excitations can be used to encode logical information in the form of  qubits. These topological qubits have robust coherence properties since they are immune to noise arising from local perturbations \cite{Kitaev2003,Kitaev2006,Nayak2008,Terhal2015}. One of the most promising candidates for topological qubits is using Majorana zero modes (MZM-s) \cite{Kitaev2001,Terhal2012, Landau2016,Roy20172}. Four MZM-s can be used to encode a qubit \cite{Bravyi2006}. The non-Abelian braiding statistics of the MZM-s, together with magic state distillation, can be used to perform all the single and two-qubit gates required for universal quantum computing  \cite{Bravyi2005,Kitaev2006,Bravyi2006, Leijnse2012}. There are several proposals to experimentally realize these MZM-s \cite{Fu2008,Alicea2010,Akhmerov2011}. One of the most promising directions is to realize experimentally Kitaev's toy model of a 1D, spinless, \textit{p}-wave superconductor \cite{Kitaev2001}.  Recently, remarkable progress has been made in experimental realizations of this model \cite{Delft2012,Copenhagen2016,Delft2017}. Alternately, MZM-s were already proposed to exist in 2D chiral \textit{p}-wave superconductors \cite{Read2000,Alicea2012}.

This work concerns the MZM-s that were predicted to arise in Kitaev's exactly
solvable honeycomb model \cite{Kitaev2006}. The model describes spins
interacting on a honeycomb lattice, where the nature of the interaction depends
on the direction of the link on the lattice. Due to the interaction, the spin
degrees of freedom fractionalize into Majorana fermions, interacting with an
emergent static $\mathbb{Z}_2$ gauge field \cite{Senthil2000}. Depending on the
choice of the interaction strength, the system is either in a gapped,
$\mathbb{Z}_2$ toric code phase with Abelian anyons or in a gapless phase.
Addition of an external magnetic field while being in the gapless phase opens a
gap in the spectrum \cite{Kitaev2006,Burnell2011}. It is then that vortex
defects of the static $\mathbb{Z}_2$ gauge field trap MZM-s, which have the
desired non-Abelian exchange statistics and can, in principle, be used for
quantum information processing. As will be shown below, the model in this phase
can be mapped to the chiral \textit{p}-wave superconductor which then naturally
gives rise to the MZM-s \cite{Burnell2011,Roy2017,Hur2017}. 

Apart from hosting MZM-s that are generally interesting for the purpose of quantum computing, the honeycomb model describes a quantum spin liquid (QSL). The latter is a phase of matter that is highly frustrated and has no ordered ground state even at zero temperature. In the last decade, there has been a lot of theoretical and experimental effort to characterize different materials with interactions similar to that of Kitaev's honeycomb model and finding a QSL phase of matter \citep{Lee2008, Jackeli2009, Leon2010, Witczak-Krempa2014, Rau2014, Winter2017, Zhou2017, Savary2017, Hermanns2018, Ducatman2018, Hickey2018}. There is experimental evidence that certain materials, among them for example $\alpha$-RuCl$_3$, are dominated by the interaction of the Kitaev honeycomb model \cite{Khaliullin2005,Jackeli2009, Plumb2014}. These materials show QSL behavior above certain temperatures. However, if cooled sufficiently, all these candidates tend to magnetically order due to additional non-Kitaev interactions.
Recently, an NMR measurement of H$_3$LiIr$_2$O$_3$ showed no sign of magnetic ordering at all \citep{Kitagawa2018} while a large set of low energy states was observed in the specific heat and the NMR measurement results. This indicates spin liquid behavior in H$_3$LiIr$_2$O$_3$ that still lacks a good model describing the findings.
As a reaction to this experiment, different proposals are under discussion to explain the results\citep{Slagel2018,Kimchi2018,Knolle2018}. One promising approach is to consider disorder in the model's interaction strengths, analyzing the ``bond-disordered Kitaev model" \citep{Knolle2018}. 

In this work, we treat a different problem where we analyze the non-Abelian phase of Kitaev's honeycomb model in the presence of quenched disorder of vortex configurations. Since the gauge field and its vortices are static quantities in Kitaev's model, we expect this situation to be well-described by an average over quenched disordered configuration of vortices. To characterize such a system, the dynamical structure factor is indispensable, which can potentially be measured with neutron scattering. We provide analytic results of the low energy dynamical spin structure factor. We find that in the presence of vortices, the dynamical structure factor has an additional peak centered at the energy $E_{\text{fl}}$  (also called the flux gap, which is the energy added by an excitation of the $\mathbb{Z}_2$ gauge field due to the presence of two additional vortices) with an unusual decay behavior proportional to $\ln[\hat{\omega}/(\omega-E_{\text{fl}})\ln(\hat{\omega}/(\omega-E_{\text{fl}}))]^2$, where $\omega$ is the frequency and $\hat{\omega}$ defines a scale that depends on the applied magnetic field strength. This peak and the decay is a signature of the vortices present in the sample. Moreover, we provide analytical results for the energy-splitting due to hybridization of MZM-s and the wave functions, which agree well with previous numerical findings\cite{Nussinov2008,Cheng2009,Lahtinen2011,Knolle2014,Knolle2014, Gohlke2017}. 

The paper is organized as follows. In Sec.~\ref{Sec.:System}, we describe Kitaev's honeycomb model in a magnetic field in terms of Majorana fermions by using a Jordan-Wigner transformation. Subsequently, we provide a continuum description of the system. In Sec.~\ref{Sec.:WaveFunction}, we calculate the wave functions of the MZM-s in the continuum model. In Sec.~\ref{Sec.:EnergySplitting}, we present results for the splitting of the ground state energy due to a finite overlap of the wave function of two MZM-s. In Sec.~\ref{Sec.:Correlator}, we calculate the low energy contribution to the dynamical structure factor in presence of two vortices. Finally, in Sec.~\ref{Sec.:DisorderAverage} we consider a quenched disordered distribution of vortices and calculate the dynamical structure factor. In Sec~\ref{Sec.:Conclusion}., we summarize our findings. 
\section{Description of the Model Hamiltonian}
\label{Sec.:System}
\begin{figure}
\includegraphics[scale=1.0]{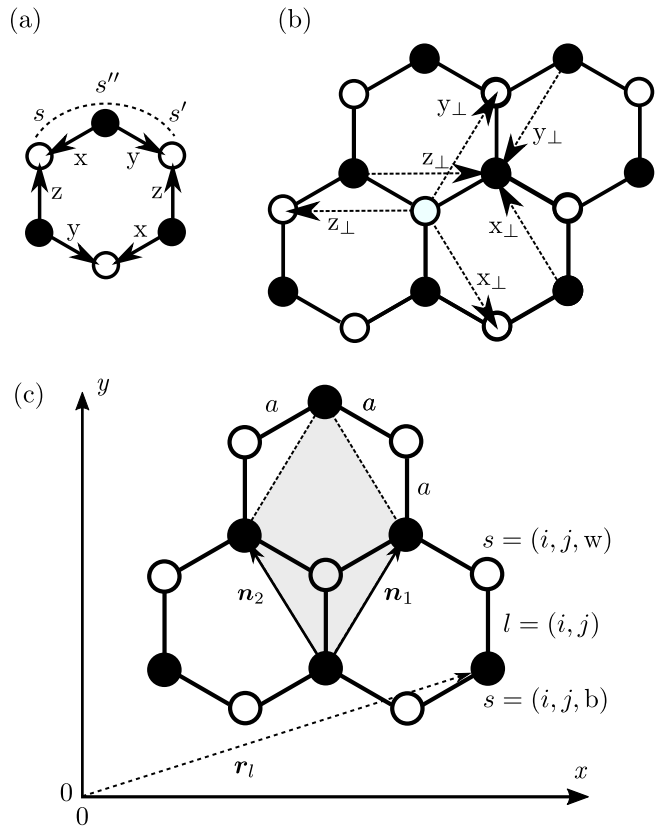}
\caption{\label{Fig.:System}(a) A plaquette of the honeycomb lattice. The letters x,y and z label the type of the link between sites and indicate if the corresponding coupling in the Hamiltonian $H_0$ is of a xx, yy or zz-type. The direction of the links represents an ordering of the operators in the Hamiltonian. They do not have a meaning for the spin Hamiltonian in Eq.~(\ref{Eq.:Ham0}), but are important for the Majorana hopping Hamiltonian in Eq.~(\ref{Eq.:HoppingHam}). The three labels $s$, $s'$ and $s''$ represent an example configuration that contributes to the magnetic perturbation $H_K$. The dashed line indicates how the three neighboring sites partly encircle a plaquette as mentioned in the main text. (b) The plot shows the next to nearest neighbor links (nnn-links) $x_\perp$, $y_\perp$ and $z_\perp$ for a site on the black and one on the white sub-lattice. These links are perpendicular to the original links and point in opposite directions for the white and the black sub-lattices. The nnn-links describe the hopping in $H_M$ of Eq.~(\ref{Eq.:HoppingHam}).  (c) The honeycomb lattice with the choice of the unit-cell (gray) and the basis vectors $\bm{n}_1$ and $\bm{n}_2$. Each unit-cell contains a white and black dot connected with a vertical z-link. The lattice constant between sites is $a$.}
\end{figure}
\begin{figure}
\includegraphics[scale=1.0]{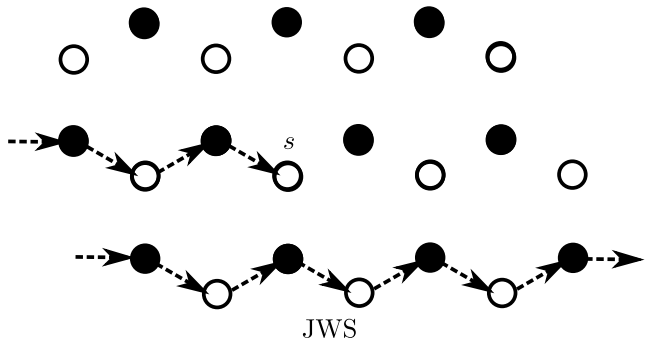}
\caption{\label{Fig.:JordanWigner}
The dashed line shows the Jordan-Wigner string (JWS) used in Eq.~(\ref{Eq.:JordanWigner}). It runs in zig-zag along a horizontal line from the left to the right. It jumps from the right end of the system back to the left where it runs again from left to right on the horizontal line above, until it reaches the site $s$.}
\end{figure}

We consider spins on a honeycomb lattice, in the presence of an external magnetic field. The total system Hamiltonian is given by $H=H_0+H_M$.
The lattice consists of two sublattices as indicated by the black and white sites in Fig.~\ref{Fig.:System}.  Every site is labeled by the index $s=(i,j,b/w)$, where $i,j$ label the position of a z-link [see the labels of the links in Fig.~\ref{Fig.:System} (a)] while $b$ or $w$ indicates if the site sits on the black or the white sublattice. The Hamiltonian $H_0$ reads
\begin{align}
\label{Eq.:Ham0}
H_0=&-J_x\sum_{\text{x-links}}\sigma_{s}^x\sigma_{s'}^x-J_y\sum_{\text{y-links}}\sigma_{s}^y\sigma_{s'}^y\nonumber\\
&-J_z\sum_{\text{z-links}}\sigma_{s}^z\sigma_{s'}^z,
\end{align}
where the sum over x-links runs over contributions where $s$ and $s'$ are connected by a link with label x in Fig.~\ref{Fig.:System} (a) (and equivalent for y and z). At this point, the arrows in the figure are irrelevant, but they will be important further below. In the absence of a magnetic field, the system can either be in a phase with a gapped spectrum or in a phase with gapless spectrum, depending on the choice of couplings. Following Ref. [\onlinecite{Kitaev2006}], we analyze the system at its isotropic point $J_x=J_y=J_z=J$ that corresponds to the gapless phase. Applying an additional magnetic field to the system breaks the time-reversal invariance. This perturbation opens a gap in the spectrum and then, the $\mathbb{Z}_2$ vortex excitations trap MZM-s. The additional contribution to the Hamiltonian due to the magnetic field is given by 
\begin{align}
\label{Eq.:MagHam}
H_M=&-K\sum_{\langle s,s',s'' \rangle}\sigma_{s'}^x\sigma_{s}^y\sigma_{s''}^z,
\end{align}
with $K \ll J$. The sum over $\langle s,s',s'' \rangle$ denotes a sum over three neighboring sites for which we have to satisfy the following rules when assigning  $s,s',s'' $ to the sites. To understand these rules, note that the three neighboring sites partly encircle a plaquette as shown in Fig.~\ref{Fig.:System}(a).  The label $s'$ and thus the $\sigma_x$ operator always belongs to the site that has an x-link pointing away from this partly encircled plaquette. The same applies for the $\sigma_y$ operator with a y-link and the $\sigma_z$ operator with a z-link. An example configuration is shown in Fig.~\ref{Fig.:System}(a). 
Next, we show how the Hamiltonian of the system can be mapped onto that of non-interacting fermions. To this end, we use the Jordan-Wigner transform\cite{Nussinov2008}
\begin{align}
\label{Eq.:JordanWigner}
\sigma_{s}^{+}&=\biggl(\prod_{s'\in \text{JWS}}\sigma_{s'}^z\biggr)c_{s}^\dagger,\\
\sigma_{s}^z&=2 c_{s}^\dagger c_{s}-1,
\end{align}
where the product symbol denotes a product over all sites along the Jordan-Wigner string (JWS) from its beginning to $s$, along the path shown in Fig.~\ref{Fig.:JordanWigner}. Moreover, we introduce the Majorana operators
\begin{align}
A_{s}=i(c_{s}^\dagger-c_{s}), B_{s}=(c_{s}^\dagger+c_{s}), \hspace{10px}\text{if}\; s=(i,j,w),\nonumber\\
 A_{s}=(c_{s}^\dagger+c_{s}),B_{s}=i(c_{s}^\dagger-c_{s}), \hspace{10px}\text{if}\; s=(i,j,b).
\end{align}
The original spin Hamiltonian maps to a Hamiltonian that is quadratic in the Majorana operators with only nearest neighbor interactions. In addition, the magnetic perturbation provides a coupling between next-to-nearest neighbors. The Hamiltonians are given by
\begin{align}
\label{Eq.:HoppingHam}
H_0=&-i J\biggl[\sum_{\text{x-links}} A_{s} A_{s'}+\sum_{\text{y-links}}A_{s} A_{s'},\nonumber\\
&+\sum_{\text{z-links}} \alpha_l A_{s}A_{s'}\biggr],\nonumber\\
H_M=&K\biggl[\sum_{\text{x}_{\perp}\text{-link}}\alpha_{l}A_{s}A_{s'}+\sum_{\text{y}_{\perp}\text{-link}}\alpha_{l'}A_{s}A_{s'}\nonumber\\
&\hspace{4px}+\sum_{\text{z}_{\perp}\text{-link}}A_{s}A_{s'}\biggr],
\end{align}
where the terms in $H_0$ are identical to the terms in Eq.~(\ref{Eq.:Ham0}) with the important difference that the direction of the arrows in Fig.~\ref{Fig.:System} (b) are now relevant as it defines the ordering of the product of Majorana operators. In particular, the operator $A_{s'}$ has to be to the right of $A_s$ if there is an arrow starting at the site $s'$ and terminating at $s$. The terms in $H_M$ comprise the next to nearest neighbor links labeled by x$_\perp$, y$_\perp$ and z$_\perp$ (perpendicular to the original links) as shown in Fig.~\ref{Fig.:System} (b). Note that all hopping terms are among the $A$-Majoranas while the $B$-Majoranas only appear in form of the $\hat{\alpha}_{l}=iB_{s}B_{s'}$; here $s=(i,j,w)$ denotes a site on the white, $s'=(i,j,b)$ on the black sub-lattice, and $l=(i,j)$ labels the z-link at position $(i,j)$ between the sites $s$ and $s'$. Every term in Eq.~(\ref{Eq.:HoppingHam}) that has a vertical component along a z-link is multiplied by the $\hat{\alpha}_{l}$ of the corresponding link. Since $\hat{\alpha}^2_{l}=1$, all $\hat{\alpha}_l$-s have eigenvalues $\alpha_{l}=\pm 1$. Furthermore, all $\hat{\alpha}_{l}$-s commute with the Hamiltonian and, thus, they are conserved. Denoting the plaquette by the index of the vertical link to the left of the plaquette, we find that for each plaquette, there is a conserved quantity $\hat{W}_l = \hat\alpha_{l=(i,j)}\hat\alpha_{l'=(i+1,j+1)}$ with eigenvalues $W_l=\pm1$. If $W_l=-1$, the plaquette carries a vortex, which is an excitation of the $\mathbb{Z}_2$ gauge field. Since the $W_l$-s  are conserved, it follows that the $\mathbb{Z}_2$ gauge field has no dynamics \cite{Kitaev2006, Gohlke2017}. The position and number of vortices are the physical properties that define the sector of the Hamiltonian when choosing the configuration of the  $\mathbb{Z}_2$ gauge field \footnote{Note that in the original Kitaev paper \cite{Kitaev2006} a $\mathbb{Z}_2$ degree of freedom was attached to ever link. In our model, only the z-links have this gauge degree. The reason is that using the Jordan-Wigner transform contains already a choice of gauge that is equivalent to choosing the gauge field along the x- and y-links to be 1.}. All choices that do not change these properties are gauge equivalent\citep{Kitaev2006,Alicea2012}. It turns out that each vortex hosts a localized MZM. In the following, we quantify the spatial distribution of these zero modes and their physical implications. 
 
 First, we embed the system into real space by choosing a unit-cell as shown in Fig.~\ref{Fig.:System} (c) with vectors $\bm{n}_1=a(\sqrt{3}/2,3/2)^T$ and $\bm{n}_2=a(-\sqrt{3}/2,3/2)^T$ and the lattice constant $a$. The position of the black site in each unit cells is given by the vector $\bm{r}_l=i\bm{n}_1+j\bm{n}_2$. The ground state of the system is vortex free \cite{Kitaev2006}. The simplest choice of gauge for the vortex free sector is to choose $\alpha_l=1$ for all $l$. Since this choice is translation invariant, we analyze the resulting Hamiltonian in Fourier space. The Hamiltonian is given by 
\begin{align}
H=\sum_{\bm{q}}
\begin{pmatrix}
A_w^{\bm{q}}\\A_b^{\bm{q}}
\end{pmatrix}
\begin{pmatrix}
\epsilon_q & -\kappa^*_{\bm{q}} \\ 
-\kappa_{\bm{q}} & - \epsilon_q 
\end{pmatrix}
\begin{pmatrix}
A_w^{-\bm{q}}\\A_b^{-\bm{q}}
\end{pmatrix},
\end{align}
with 
\begin{align}
A_{w/b}^{\bm{q}}=\frac{1}{\mathcal{N}}\sum_{i,j} e^{i\bm{q}\cdot\bm{r}_l}A_{s=(i,j,w/b)},
\end{align} 
where $\mathcal{N}=L^2/3a^2$ is the total number of unit-cells and $L$ the length of the system. The matrix elements read
\begin{align}
\epsilon_q/K=&\sin[\bm{q}\cdot(\bm{n}_1-\bm{n}_2)]-\sin(2\bm{q}\cdot\bm{n}_1)-\sin(2\bm{q}\cdot\bm{n}_2)\nonumber,\\
\kappa_{\bm{q}}=&\frac{i}{2}[J_x e^{-i\bm{q}\cdot\bm{n}_1 }+J_y e^{-i\bm{q}\cdot\bm{n}_2}-J_z]e^{i\bm{q}\cdot(\bm{n}_1+\bm{n}_2)/3}.
\end{align}
The off-diagonal elements $\kappa_{\bm{q}}$ in the Hamiltonian originate from $H_0$ while the diagonal elements are due to the magnetic field. The function $\kappa_{\bm{q}}$ vanishes at the Dirac points $\pm\bm{q}_D=(2\pi/3\sqrt{3},0)/a$. Close to these points the spectrum is linear. Expanding $\kappa_{\bm{q}}$ around $\pm\bm{q}_D$ leads to $\kappa_{\bm{q}}\approx(3/4)[(q_x-q_{Dx})\pm i(q_y-q_{Dy})]$.

The magnetic contribution to the Hamiltonian gives rise to a gap in the spectrum of size $E_g=4|\epsilon_{\bm{q}_D}|=2\sqrt{3}K$. As we are interested in the low energy properties of the system, we linearize the Hamiltonian in the vicinity of the Dirac points and apply a continuum approximation. For this purpose we introduce the new complex field operators $A_{w/b}(\bm{r})$. The new fermions satisfy the relation (valid in the limit $a\rightarrow 0$)
\begin{align}
\label{Eq.:NewFermions}
A_{s}/\sqrt{3}a=&e^{i\bm{q}_D\cdot\bm{r}}A_{w/b}(\bm{r})+e^{-i\bm{q}_D\cdot\bm{r}}A_{w/b}^{\dagger}(\bm{r}),\nonumber\\
\end{align}
where the factor $\sqrt{3}a$ is the square root of the area of the unit cell. The vector $\bm{r}$ without the subscript $l$ indicates that it describes a point in the continuum instead of the lattice sites $\bm{r}_l$. In this picture, the annihilation operator $A_{w/b}(\bm{r})$ corresponds to the contribution of the Dirac cone at $\bm{q}_D$ while the creation operator $A_{w/b}^{\dagger}(\bm{r})$ corresponds to the opposite Dirac cone at $-\bm{q}_D$. The new operators $A_{w/bD}(\bm{r})$ obey the canonical fermionic relations 
\begin{align}
\{A_{b/w}(\bm{r}),A_{b/w}^{\dagger}(\bm{r}')\}&=\delta^{(2)}(\bm{r}-\bm{r}'),\nonumber\\
 \{A_{b/w}^{\dagger}(\bm{r}),A_{b/w}^{\dagger}(\bm{r}')\}&=\{A_{b/w}(\bm{r}),A_{b/w}(\bm{r}')\}=0.
\end{align}
In the following, we make use of a 4D Bogoliubov-de Gennes representation and introduce the spinor $\bm{A}(\bm{r})=[A_{w}(\bm{r}),A_{b}(\bm{r}),A_{b}^\dagger(\bm{r}),A_{w}^\dagger(\bm{r})]^T$. The first two entries contain the hole (annihilation) operators and the last two entries the particle (creation) operators  \footnote{Note that in our case particles correspond to the first Dirac point while holes correspond to the other Dirac point}. Note that the operator $\bm{A}(\bm{r})$ obeys the symmetry
\begin{align}
\label{Eq.:ParticleHoleSymmetry}
[\sigma_x \tau_{x}\bm{A}(\bm{r})]^T=\bm{A}^{\dagger}(\bm{r});
\end{align} 
here $\sigma_x$ swaps the $b$/$w$ degrees of freedom while the matrix $\tau_{x}$ acts on the particle-hole degrees of freedom. In the continuum limit, the Hamiltonian reads
\begin{figure}
\includegraphics[scale=1.0]{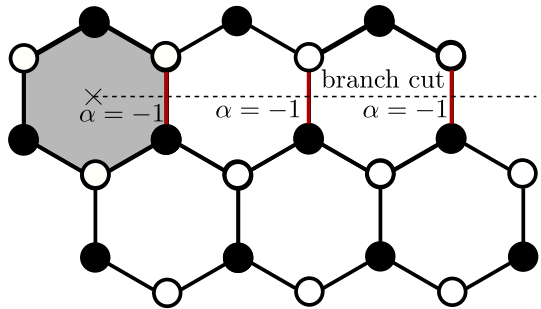}
\caption{
\label{Fig.:FieldSystem} A configuration with a vortex in the middle of the shaded plaquette marked with the cross. We choose the following gauge to describe a vortex: all $\alpha_l$ on the dashed line are set to $-1$ while all other $\alpha_l$ are kept to be $1$. In a continuum model, the dashed line will manifest itself as a branch cut. The wave function of bound Majorana modes changes sign across the branch cut.}
\end{figure}
\begin{align}
\label{Eq.:ContinuumHamiltonian}
H=&\int d^2r\bm{A}^{\dagger}(\bm{r})\mathcal{H}(\bm{r})\bm{A}(\bm{r}),
\end{align}
with the Bogoliubov-de Gennes Hamiltonian
\begin{align}
\label{Eq.:ContinuumHamiltonianDensity}
\mathcal{H}=&[E_g\sigma_z/4+iv\sigma_y\partial_{x}+iv\sigma_x\partial_{y}]\tau_0;
\end{align} 
here, we have introduced the velocity \[v=3Ja/4\] of the Majorana modes. The Hamiltonian in Eq.~(\ref{Eq.:ContinuumHamiltonianDensity}) is block diagonal in the particle-hole degrees of freedom\footnote{Note, that in the upper block the spin-up like state corresponds to a Majorana operator attached to a white site in the lattice and the spin-down like state to a Majorana operator corresponding to a black site, while in the lower block this choice is reversed.}, which is expressed by the identity operator $\tau_0$. This property originates from the fact that we ignored fast oscillating terms proportional to $e^{\pm i2\bm{q_D}\cdot\bm{r}}$.

\section{Calculation of the zero-mode wave function}
\label{Sec.:WaveFunction}
The aim of this section is to derive an analytical expression for the wave function of the bound zero modes of a sufficiently isolated vortex. The vortex is described by setting all $\alpha_l=1$ except for the $\alpha_l$ on the z-bonds along a horizontal line that starts at the vortex position and ends at infinity as shown in Fig.~\ref{Fig.:FieldSystem}. This can be expressed as a Hamiltonian $\mathcal{H}_1=\mathcal{H}+\mathcal{V}$ describing a system with a single vortex. The potential term $\mathcal{V}$ adds a vortex to $\mathcal{H}$ at the origin. As the system obeys translation invariance, the resulting zero mode is general and can later be shifted to any position in the sample. In the continuum approximation, the potential $\mathcal{V}$ changes $\mathcal{H}$ only along a horizontal line and can be implemented by changing the boundary conditions of the solution along this line: in particular, we require the wave function to change sign when crossing the line. Further below, we show that this line manifests itself as a branch cut.

For the calculation of the zero mode, we need to solve the equation 
\begin{align}
\label{Eq.:MainZeroMode}
\mathcal{H}_1\bm{\chi}=0.
\end{align}
The Hamiltonain $\mathcal{H}_1$ is block diagonal in the space of the two Dirac cones and can be solved separately for each block. Thus, the problem reduces to solving the $2\times 2$ equation
\begin{align}
\label{Eq.:MatrixFormHam1}
\begin{pmatrix}
\epsilon & \partial_{x}+i\partial_{y}\\ 
-\partial_{x}+i\partial_{y} & - \epsilon \\
\end{pmatrix}
\begin{pmatrix}
\chi_{1} \\ 
\chi_{2} \\
\end{pmatrix}=0,
\end{align}
where we have introduced $\epsilon=E_g/3Ja$, which will turn out to be the inverse decay length of the zero mode.
Due to the radial symmetry, it is simpler to solve this problem in polar coordinates. Transforming Eq.~(\ref{Eq.:MatrixFormHam1}) yields
\begin{align}
\label{Eq.:ZeroModeEquation}
\begin{pmatrix}
\epsilon & e^{i \varphi}[\partial_{r}-i\partial_{\varphi}/r)]\\
e^{-i \varphi}[\partial_{r}+i\partial_{\varphi}/r]& - \epsilon \\
\end{pmatrix}
\begin{pmatrix}
\chi_{1} \\ 
\chi_{2} \\
\end{pmatrix}=0,
\end{align}
where  $\varphi\in[0,2\pi[$ is the angle with respect to the positive $x$-axis. 
This choice implements a branch cut on the positive $x$-axis so that the ansatz
\begin{align}
\chi_1=g(r)e^{-i (n-1/2) \varphi},\nonumber\\
\chi_2=g(r)e^{-i (n+1/2) \varphi}
\end{align} with $n\in\mathbb{Z}$ satisfies the boundary conditions and changes sign when crossing the branch cut. Inserting the ansatz into Eq.~(\ref{Eq.:ZeroModeEquation}) leads to the two equations
\begin{align}
\label{Eq.:WaveFunctionEq}
n=0,\nonumber\\
(\epsilon+\partial_{r}+\frac{1}{2r})g(r)=0.
\end{align}
The first equation tells us, that the zero mode is radial symmetric. From the second equation, we find the explicit radial wave function
\begin{align}
g(r)\propto
\frac{e^{-\epsilon r}}{r^{1/2}}.
\end{align}
Note that our result is qualitatively different from the wave function that
would describe the MZM in a chiral \textit{p}-wave superconductor
\cite{Read2000,Alicea2012} even though after Jordan-Wigner transform, both
Hamiltonians are the same modulo irrelevant constants. The wave function in the case of a \textit{p}-wave superconductor with the same Hamiltonian has only an exponential decay without the power-law dependence. The difference is due to the fact that in the superconducting case, one searches for a solution which is periodic in $\varphi$ and thus, half-integer $n$-s. This leads to an edge mode around the defect which is not at zero energy. The zero-mode then arises by introducing a half-a-flux quantum magnetic vortex in the defect, which then give rise to a zero-energy mode. In the case of the honeycomb model, the vortex defect of the $\mathbb{Z}_2$ gauge field is sufficient to give rise to the zero energy mode by itself. The key point here is that the charge degree of freedom of electrons couples to electromagnetic fields. This gives rise to qualitatively different physics compared to that arising from interacting spins, even though the Hamiltonians in the two cases appear to be similar.

The lower block of Eq.~(\ref{Eq.:MainZeroMode}), corresponding to the opposite Dirac cone, has the same solution for the wave function. To satisfy the symmetry, Eq.~(\ref{Eq.:ParticleHoleSymmetry}), of the spinor $\bm{A}(\bm{r})$ the full zero mode has to combine both Dirac points. Additionally, we require the absolute value of the black and white components of the wave function to be continuous when passing through the vortex along the line of the branch cut. Using these criteria leads to the normalized spinor $\bm{\chi}(\bm{r})$ for a zero mode bound to a vortex at the origin, reading
\begin{align}
\label{Eq.:WaveFunctionSpinor0}
\bm{\chi}(\bm{r})=
\begin{pmatrix}
e^{i(\varphi/2+\pi/4)}  \\ 
e^{-i(\varphi/2-\pi/4)} \\
  e^{i(\varphi/2-\pi/4)} \\ 
 e^{-i(\varphi/2+\pi/4)} \\
\end{pmatrix}g(r),
\end{align}
where $g(r)=(\epsilon/4\pi r)^{1/2}e^{-\epsilon r}$. The wave function decays exponentially due to the decay length $\epsilon^{-1}$ introduced by the magnetic field in addition to an algebraic decay $\propto r^{-1/2}$. The latter corresponds to the conventional decay of a radial symmetric wave. 

\begin{figure}
\includegraphics[scale=1.0]{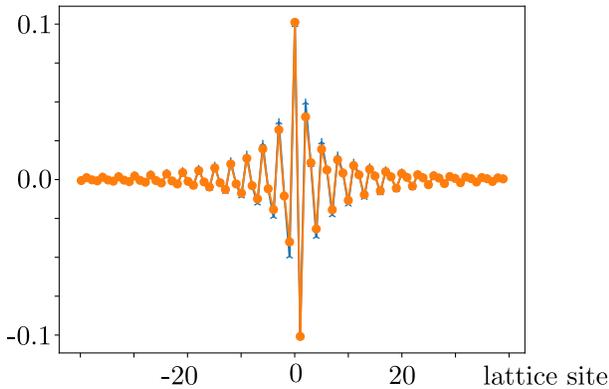}
\caption{
\label{Fig.:WaveFunction} The plot shows the analytic result for the wave function of the zero mode (dots) on top of the numerical exact wave function (little crosses) obtained by numerical diagonalization of Eq.~(\ref{Eq.:HoppingHam}). The wave functions are evaluated along the branch cut on the white sublattice (with $K=0.03J$, for a vortex at the origin). The dots and crosses represent the wave function on lattice sites while the lines connecting the dots are a guide to the eye. The plot shows that the analytic results are valid even for relatively small distances from the vortex core. }
\end{figure}

From Eq.~(\ref{Eq.:NewFermions}), we can read off the transformation behavior of the wave function when we shift the Hamiltonian by a vector $\bm{R}_k$, representing the position of a vortex not at the origin. The hole components are multiplied by a phase factor $e^{-i\bm{q_D}\cdot\bm{R}_k}$, while the particle components need to be multiplied by $e^{i\bm{q_D}\cdot\bm{R}_k}$. This leads to the general MZM wave function bound to a vortex at position $\bm{R}_k$ given by
\begin{align}
\label{Eq.:WaveFunctionSpinor}
\bm{\chi}_{k}(\bm{r})=
\begin{pmatrix}
e^{i(\varphi_k/2-\bm{q_D}\cdot\bm{R}_k+\pi/4)}  \\ 
e^{-i(\varphi_k/2+\bm{q_D}\cdot\bm{R}_k-\pi/4)} \\
  e^{i(\varphi_k/2+\bm{q_D}\cdot\bm{R}_k-\pi/4)} \\ 
 e^{-i(\varphi_k/2-\bm{q_D}\cdot\bm{R}_k+\pi/4)} \\
\end{pmatrix}g(|\bm{r}-\bm{R}_k|),
\end{align}
where $\varphi_k\in[0,2\pi[$ is the angle defined with respect to an axis parallel to the $x$-axis that runs through the vortex core at $\bm{R}_k$. From this we find the zero mode operator 
\begin{align}
\label{Eq.:MajoranaOperator}
\gamma_k=&\int d^2r e^{-\epsilon |\bm{r}-\bm{R}_k|}(\epsilon/4\pi |\bm{r}-\bm{R}_k|)^{1/2} \nonumber\\
&\times[f_w A_{w}(\bm{r})+f_w^* A^{\dagger}_{w}(\bm{r})+f_bA_{b}(\bm{r})+f_b^* A^{\dagger}_{b}(\bm{r})],
\end{align}
with $f_w=e^{i(\varphi_k/2-\bm{q_D}\cdot\bm{R}_k+\pi/4)}$ and $f_b=e^{-i(\varphi_k/2+\bm{q_D}\cdot\bm{R}_k-\pi/4)}$. Note that $\gamma_k$ is manifestly Hermitian and therefore, represents a MZM.
In Fig.~\ref{Fig.:WaveFunction} we compare the numerical exact wave function, obtained by numerically diagonalizing Eq.~(\ref{Eq.:HoppingHam}), to our analytic expression. The agreement is rather good even for moderate distances from the vortex core. 
\section{Computation of the energy splitting in the presence of two vortices}
\label{Sec.:EnergySplitting}
\begin{figure}
\includegraphics[scale=1.0]{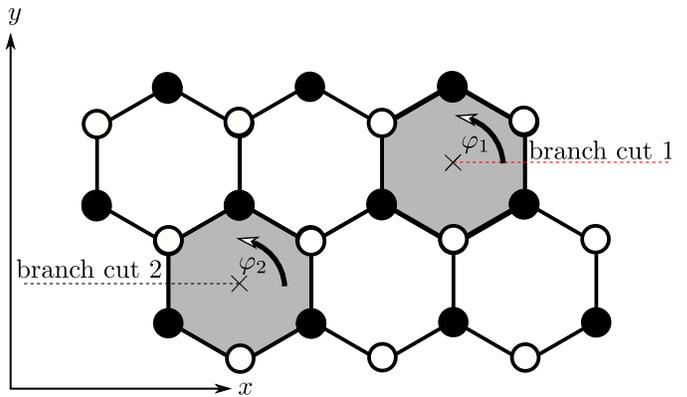}
\caption{
\label{Fig.:VortexGeometry} The sketch shows the geometry we choose for the calculation of the energy splitting due to hybridization of two MZM-s. The gray plaquettes with a cross mark the position of a vortex. The angle $\varphi_1 $ of the vortex further right on the $x$-axis is defined with respect to an axis that runs through the vortex parallel to the $x$-axis such that $\varphi_1\in[0,2\pi[$. This adds a branch cut for the wave function pointing along the $x$-axis to positive infinity. The vortex further left on the $x$-axis is described by the angle $\varphi_2$ with respect to an axis that runs through this vortex parallel to the $x$-axis. This time $\varphi_2\in[-\pi,\pi[$, so that the branch cut points along the $x$-axis to negative infinity. We can choose to evaluate the matrix element $\langle \bm{\chi}_2| \mathcal{H}_2 |\bm{\chi}_1\rangle$ on one of the branch cuts. In the main text, we choose to evaluate it on branch cut 1, indicated by the red color. }
\end{figure}
Above, we calculated the wave function  of the zero mode bound to a single vortex in the Kitaev honeycomb model. If there are two vortices, each of these carries a zero mode. Provided that the vortices are separated sufficiently, this leads to a two-fold degenerate ground state. However, if the vortices approach each other, the zero modes hybridize to give rise to a conventional fermionic mode and, as a result, the ground state is no longer degenerate. The aim of this section is to calculate the energy difference $\Delta$ between the ground state and the first excited state that arises due the hybridization between two MZM-s.

We split the Hamiltonian $\mathcal{H}_2=\mathcal{H}_0+\mathcal{V}_1+\mathcal{V}_2$, describing the system and the two vortices, into the Hamiltonian $\mathcal{H}_0$ for a system without vortices and the two potential terms $\mathcal{V}_1$ and $\mathcal{V}_2$, which each add a vortex. We project the eigenvalue equation $\mathcal{H}_2\Psi=E \Psi$ into the low energy sub-space of the two zero modes. To second order, the energy splitting is given by
\begin{align}
\Delta= 2|\langle \bm{\chi}_2| \mathcal{H}_2 |\bm{\chi}_1\rangle |,
\end{align}
where $\bm{\chi}_{1/2}$ are the MZM wave functions for a vortex at position $\bm{R}_{1/2}$ as given by Eq.~(\ref{Eq.:WaveFunctionSpinor}). Our main task is to determine the matrix element $\langle \bm{\chi}_2| \mathcal{H}_2 |\bm{\chi}_1\rangle$. For the evaluation of the latter we can use the fact that we are dealing with zero modes with the properties $(\mathcal{H}_0+\mathcal{V}_1)\bm{\chi}_1=(\mathcal{H}_0+\mathcal{V}_2)\bm{\chi}_2=0$
and conclude
\begin{align}
\label{Eq.:MatrixElement}
\langle \bm{\chi}_2| \mathcal{H}_2 |\bm{\chi}_1\rangle=\langle \bm{\chi}_2| \mathcal{V}_1|\bm{\chi}_1\rangle=\langle \bm{\chi}_2| \mathcal{V}_2|\bm{\chi}_1\rangle.
\end{align}
In Sec.~\ref{Sec.:WaveFunction}, we implemented the potential that adds a vortex by changing the boundary condition of the unperturbed system for the wave function. However, here it is more convenient to have an explicit expression for the vortex potentials $\mathcal{V}_1$ and $\mathcal{V}_2$. The potentials needs to switch the sign for all hopping interactions along the branch cut of the corresponding vortex. For the calculation of the energy splitting, we introduce the geometry as shown in Fig.~\ref{Fig.:VortexGeometry}. We choose the branch cut of the vortex further right on the $x$-axis to point to positive infinity. Thus the angle around the vortex $\varphi_1$ with respect to an axis that runs through the vortex parallel to the $x$-axis is defined as $\varphi_1\in[0,2\pi[$. For the vortex further left on the $x$-axis, we choose the branch cut to point in the opposite direction so that the angle $\varphi_2$ with respect to an axis that runs through this vortex parallel to the $x$-axis is defined as $\varphi_2\in[-\pi,\pi[$. This choice keeps the two potentials locally separated and avoids that $\mathcal{V}_1$ and $\mathcal{V}_2$ have terms in common. A simple way to find the explicit form of the potentials $\mathcal{V}_1$ and $\mathcal{V}_2$ is to consider twice the negative $z$-interaction terms along the branch cut of the second quantized Hamiltonian. By applying the continuum approximation and using the Bogoliubov-de Gennes representation to
\begin{align}
V_{1}=&-2iJ\sum_{\text{b}_{1}} A_{s=(i,j,w)}A_{s'=(i,j,b)}\nonumber\\
=& \int d^2r \bm{A}^{\dagger}(\bm{r})\mathcal{V}_{1}(\bm{r})\bm{A}(\bm{r}),
\end{align}
we can read of the potential $\mathcal{V}_1$. Here we defined $\text{b}_{1}$ as the set of z-links on the branch cut that belongs to the vortex with branch cut 1.
The potential assumes the form
\begin{align}
\mathcal{V}_{1}(\bm{r})=&-4i v \Theta[x-(\bm{R}_{1})_x]\delta[y-(\bm{R}_{1})_y]\nonumber\\ 
&\times\begin{pmatrix}
i\sigma_y &  0\\ 
0  &-i\sigma_y \\
\end{pmatrix},
\end{align} 
where we ignored the fast oscillating terms $\propto e^{\pm 2 i \bm{q}_D\cdot\bm{r}}$. Here, $(\bm{R}_{1})_x$ is the $x$-component and $(\bm{R}_{1})_y$ the $y$-component of the vortex position $\bm{R}_{1}$. In this continuum approximation, the potential has finite weight only on the branch cut, which is ensured by the $\Theta$-function in $x$-direction and the $\delta$-distribution in $y$-direction. The $\delta$-distribution is multiplied by a factor of $3a$, which is the width of the unit-cell in $y$-direction. By acting with the potential on the wave function of the MZM, it adds a minus sign to the black components of the wave function, while leaving the white components untouched. This implements the desired sign change of the wave function across the branch cut, as the white components sit above while the black components sit below the branch cut.

Proceeding similarly, we can obtain the potential $\mathcal{V}_2$. However, for the following calculation we choose to evaluate the matrix element $\langle\bm{\chi}_2| \mathcal{V}_1 |\bm{\chi}_1\rangle$, which contributes only on the branch cut 1 (see Fig.~\ref{Fig.:VortexGeometry}). This leads to the ambiguity that both, the choice $\varphi_1=0$ and $\varphi_1=2\pi$, represent the wave function on the branch cut. However, if we consider the real lattice, we need to take into account again that sites on the black sub-lattice always sit below the branch cut while sites on the white sub-lattice always sit above the branch cut. To implement this fact in the continuum model, we have to set $\varphi_1=0$ for the spinor components of the white sub-lattice and $\varphi_1=2\pi$ for the spinor components of the black sub-lattice. Evaluating the matrix element,
we find 
\begin{align}
\langle\bm{\chi}_2| \mathcal{V}_1 |\bm{\chi}_1\rangle =&-16iv\cos(\bm{q_D}\cdot\bm{R})\nonumber\\
&\hspace{-5pt}\times\int_{(\bm{R}_{1})_x}^{\infty}dx\cos(\frac{\varphi_2}{2})g(|\bm{r}-\bm{R}_1|)g(|\bm{r}-\bm{R}_2|),
\end{align} 
where $\bm{R}=\bm{R}_1-\bm{R}_2$. This term has a direction dependence by the factor $\cos(\bm{q}_D\cdot\bm{R})$ that implements the properties of the honeycomb lattice in the continuum model. Beyond that, we expect the splitting to be isotropic because we treat the isotropic Kitaev model. Therefore, the result of the integral needs to be independent of the angle $\varphi_2$. We checked analytically that the integral, except of the factor $\cos(\bm{q}_D\cdot\bm{R})$, indeed does not depend on the angle $\varphi_2$, as long as the distance $|\bm{R}|$ between the vortices is fixed. Thus, we can choose $\varphi_2=0$ for simplicity and define $\tilde{x}=x-R/2$ to find
\begin{align}
\langle\bm{\chi}_2| \mathcal{V}_1 |\bm{\chi}_1\rangle =&-4iv\cos(\bm{q_D}\cdot\bm{R})\frac{\epsilon }{\pi}e^{-\epsilon R}\int_{0}^{\infty}dx \frac{\epsilon e^{-2\epsilon \tilde{x}}}{\sqrt{\tilde{x}(\tilde{x}+R)}}\nonumber\\
=&-4iv\cos(\bm{q_D}\cdot\bm{R})\sqrt{\frac{\epsilon}{2\pi R}}e^{-\epsilon R},
\end{align}
where we assumed large distances between the vortices with $\epsilon R\gg 1$ and replaced $\tilde{x}+R\approx R$, valid in regions where the integrand has relevant contributions. 
With this, the energy splitting reads
\begin{align}
\label{Eq.:Splitting}
\Delta= 8v|\cos(\bm{q_D}\cdot\bm{R})|\sqrt{\frac{\epsilon}{2\pi R}}e^{-\epsilon R}.
\end{align}
The splitting has the same decay properties as the Majorana wave functions. The only angular dependency comes in by the sixfold symmetric factor $|\cos(\bm{q_D}\cdot\bm{R})|$. Note that there are no vortex positions on the lattice where the prefactor is exactly zero. However, for large distances between the vortices almost any vector $\bm{R}$ can be approximated and the splitting oscillates depending on the direction of of the vortex separation. 
\section{Calculation of the dynamical spin-spin structure factor}
\label{Sec.:Correlator}
In Sec.~\ref{Sec.:WaveFunction} and Sec.~\ref{Sec.:EnergySplitting}, we have calculated the MZM wave functions and the splitting of the ground state energy due to finite overlap of two such MZM-s. In this section, we treat the case of a large number ($N$) of MZM-s and analytically calculate the dynamical structure factor $S_q(\omega)$ for energies below the gap using the energy splitting and MZM wave functions obtained above. We consider the case where the temperature $T$ is below the gap $E_g$ with $k_B T\ll E_g$ but still large compared to the scale of the average hybridization splitting of two Majorana modes with $k_BT\gg v\sqrt{\epsilon/\lambda}e^{-\epsilon\lambda}$ [see Eq.~\eqref{Eq.:Splitting}], where $\lambda$ is the typical distance between vortices. In this limit the system is in an equal mix of the almost-degenerate ground states. This low energy eigenspace is spanned by the $2^N$ states $|k\rangle$, satisfying $H|k\rangle=E_k|k\rangle$ with energies $E_k$. Here, $k\in\{0,\dots,2^{N/2}-1\}$ is a number, where each bit represents the occupation of an eigenmode in the low energy space. 
The structure factor is then given by 
\begin{align}
\label{Eq.:StructureFactor}
S^{ab}_{\bm{q}}(\omega)=\frac{1}{\mathcal{N}}\sum_{s,s'} e^{i \bm{q}\cdot(\bm{r}_s-\bm{r}_{s'})} \int dt e^{i\omega t}S^{ab}_{ss'}(t),
\end{align}
where $a,b\in \{x,y,z\}$ and $S^{ab}_{ss'}(t)$ reads
\begin{align}
\label{Eq.:Correlator}
S^{ab}_{ss'}(t)=&\frac{1}{2^{N/2}}\sum_{k}\langle k | \sigma^{a}_{s}(t)\sigma^{b}_{s'}(0)|k\rangle\nonumber\\
=&\frac{1}{2^{N/2}}\sum_{k}\langle k | e^{i H t}\sigma^{a}_{s} e^{-i H t}\sigma^{b}_{s'}|k\rangle\nonumber\\
=&\frac{1}{2^{N/2}}\sum_{k}\langle k | \sigma^{a}_{s} e^{-i H t}\sigma^{b}_{s'}|k\rangle e^{i E_k t}.
\end{align}
Here $\sigma^{a}_{s}(t)$ are the Pauli matrices in the Heisenberg picture. The spin-spin correlation in the Kitaev honeycomb model is ultra short ranged and only nearest neighbor and on-site correlators that correlate spins from the same type, i.e. $\sigma_a\sigma_a$, add finite contributions \cite{Kitaev2006, Tikhonov2011, Knolle2015}. Therefore, we have $S^{aa}_{ss'}\neq 0$ only if $s=s'$ or if $s$ and $s'$ represent nearest neighbors. As we are working with the isotropic model ($J_x=J_y=J_z$), the structure factor is the same for all three types of interactions and directions. Therefore, we only calculate the $zz$-structure factor and suppress the interaction labels from now on. For the $zz$-structure factor, only sites separated by a z-link and on-site terms contribute to Eq.~(\ref{Eq.:Correlator}). Thus, we can replace $\bm{r}_s-\bm{r}_{s'}=\pm a\bm{e}_y$ for the z-link contribution and $\bm{r}_s-\bm{r}_{s'}=0$ for the on-site contribution resulting in
\begin{align}
\label{Eq.:ZCorrelator}
S_{\bm{q}}(\omega)=&\frac{1}{\mathcal{N}}\sum_{s}\int dt  e^{i\omega t}\Bigl[S_{ss}(t)\nonumber\\
&+e^{i aq_y}S_{s=(i,j,w)s'=(i,j,b)}(t)\nonumber\\
&+e^{-i aq_y}S_{s'=(i,j,b)s=(i,j,w)}(t)\Bigr],
\end{align}
where $q_y$ is the y-component of the momentum vector $\bm{q}$.
Applying the Jordan-Wigner transformation leads to the expression
\begin{align}
S_{ss}(t)=&\frac{i}{2^{N/2}}\sum_{k}\langle k| A_{s} e^{-i H_+ t}A_{s}|k\rangle e^{i E_k t},
\end{align}
if $s=s'$ and
\begin{align}
S_{ss'}(t)=&-\frac{1}{2^{N/2}}\sum_{k}\langle k| A_{s} e^{-i H_+ t}A_{s'}|k\rangle e^{i E_k t,},
\end{align}
if $s\neq s'$. In these expressions, the operators $A_s$ are the Majorana operators introduced in Sec. \ref{Sec.:System}. The Hamiltonian $H_+$ that appears in the time evolution of the transformed matrix elements is the Hamiltonian $H$ with two additional vortices in the plaquettes directly adjacent to the link between sites $i$ and $j$ \cite{Knolle2015}. It satisfies the equation $H_+|l_+\rangle=E_{l_+}|l_+\rangle$ with eigenmodes $|l_+\rangle$ and energy $E_{l_+}$. The label $l$ represents a bit string like the label $k$, with $l\in\{0,\dots,2^{N/2}-1\}$, where each bit represents the occupation of a mode. The energy of the mode hosted by the two extra vortices in $H_+$ is close to the energy of the gap, because the vortices are separated only by one plaquette and thus the energy splitting is maximal. Therefore, the extra vortices are not relevant for the low energy structure factor, except for an energy shift that is caused by the raise in excitations of the gauge field from $N$ to $N+2$ vortices. This energy shift is called the flux gap $E_{\text{fl}}=E_{0_+}-E_0$, which is the difference in the ground state energies of the Hamiltonian $H$ and $H_+$ (at the isotropic point $E_{\text{fl}}\approx0.26J$, see Ref.~\onlinecite{Kitaev2006}). 

It is useful to write down the Lehman representation of the dynamical structure factor reading (for $s\neq s'$, otherwise it is multiplied by $-i$)
\begin{align}
\label{Eq.:LehmanRealSpace}
S_{ss'}(\omega)=&\int dt e^{i\omega t}S_{ss'}(t)\nonumber\\
=&-\frac{1}{2^{N/2}}\sum_{k,l_+} \langle k|  A_{s}|l_{+}\rangle\langle l_{+} | A_{s'}|k\rangle\nonumber\\
&\hspace{55pt}\times \delta[E_k-E_{l_+}+(\omega-E_{\text{fl}})].
\end{align}
We assume a dilute gas of vortices with a low vortex density satisfying $\lambda \epsilon\gg 1$. This implies that for the creation of a fermionic eigenmode it is a good approximation to take only two MZM-s localized close to each other into account. The operators that create single eigenmodes are given by the superposition of two MZM-s $\gamma_{j,A}$ and $\gamma_{j,B}$ bound to vortices at position $\bm{R}_{j,A}$ and $\bm{R}_{j,B}$ with $z^{\dagger}_j=(\gamma_{j,A}\pm i\gamma_{j,B})/2$. The pairing of the zero modes $\gamma_{j,A}$ and $\gamma_{j,B}$ has to be done in a way that the sum of the distance between all chosen pairs of vortices in the sample is minimized. The choice of the plus or minus sign has to be taken so that the operator $z_j$ annihilates the ground state $|0\rangle$, which depends in general on the sign of the energy splitting in Eq.~(\ref{Eq.:Splitting}) without the absolute value. Note that the final structure factor will not depend on these signs. The subgap states are thus given by  $|k\rangle=\prod_j (z_j^\dagger)^{k_j}|0\rangle$, where $k_j$ is the $j$th bit of $k$. As the creation and annihilation operators $z_j^{\dagger}$ and $z_j$ are localized objects, they do not differ significantly for $H$ and $H_+$, especially as $H_+$ differs only locally from $H$, too. Therefore we can assume that the eigenstates of $H_+$ are created by the same operators as the one for $H$ with $|l_+\rangle=\prod_j (z_j^\dagger)^{l_j}|0_+\rangle$, where $l_j$ is the $j$th bit of $l$.

Only matrix elements between states $|k\rangle$ and $|l_{+}\rangle$ that differ in the occupation of a single bound mode have a finite contribution to Eq.~(\ref{Eq.:LehmanRealSpace}). This stems from the fact that the Majorana operators $A_s$ change the fermion parity by one. In other words, only matrix elements labeled by $l$ and $k$ that differ by a single bit contribute. Labeling the position of this bit by $j$, one of the sums over $k$ and $l_+$ in Eq.~(\ref{Eq.:LehmanRealSpace}) can be reduce to a sum over $j\in{1,\dots,N/2}$. The other sum results in a simple multiplication by the prefactor $2^{N/2-1}$, because there are $2^{N/2-1}$ different states that differ by the same single bit of $k$ and $l$. All these states add the same contribution to Eq.~(\ref{Eq.:LehmanRealSpace}). This leads to
\begin{align}
S_{ss'}(\omega)=&-\frac{1}{2}\sum_{j=1}^{N/2} [S_{ss',j} \delta(\omega-E_{\text{fl}}-E_{j})\nonumber\\
&\hspace{25pt}+S_{s's,j} \delta(\omega-E_{\text{fl}}+E_{j})],
\end{align}
where $E_j$ is the energy of mode $z_j$ and $S_{ss',j}$ is a single mode term that we introduce as
\begin{align}
S_{ss',j}=&\langle 0 |  A_{s} z^{\dagger}_j|0_{+}\rangle\langle 0_{+}|z_j  A_{s'}|0\rangle.
\end{align}
For the calculation of this term, it is convenient to apply the continuum approximation, so that we can use the Majorana operators from Eq.~(\ref{Eq.:MajoranaOperator}) to express the fermionic modes $z_j$. By additionally expressing the Majorana operators $A_s$ with Eq.~(\ref{Eq.:NewFermions}) by the Dirac fermions $A_{w/b}(\bm{r})$, we can proceed with the calculation by using the canonical commutation relations between the Dirac fermions. 
If we first take only the on-site contributions $S_{ss,j}$ into account, we find in the continuum approximation 
\begin{align}
\label{Eq.:OnSiteContribution}
S_{ss,j}(\bm{r})=&2\zeta\sum_{f\in \{A,B\}} g^2_{j,f}\{1+\sin[\bm{q}_D\cdot(\bm{r}-\bm{R}_{j,f})+\varphi_{j,f}]\},
\end{align}
where the continuous vector $\bm{r}$ points to the position of the unit cell containing the site $s$. In the continuum approximation, the label $ss$ is kept only for the purpose of indicating that $S_{ss,j}(\bm{r})$ represents the on-site contribution. The function $g_{j,f}=g(|\bm{r}-\bm{R}_{j,f}|)$ is the decaying prefactor of the spinor given in Eq.~\eqref{Eq.:WaveFunctionSpinor} and $\zeta$ is the overlap between the vacua of the different flux sectors with approximately $\zeta=|\langle 0_{+}|0\rangle|^2\approx0.8$ at the isotropic point of the Kitaev honeycomb model \cite{Knolle2015}. The angles $\varphi_{j,A}$ and $\varphi_{j,B}$ are the angles corresponding to the vortices, which are defined with respect to an axis that runs through the vortex parallel to the $x$-axis as shown in Fig.~\ref{Fig.:VortexGeometry}. The on-site contribution consists of terms that depend only on the wave function of a single MZM. In contrast, the nearest neighbor contribution to the dynamical structure factor $S_{s=(i,j,w)s'=(i,j,b),j}$ is complex with imaginary and real part
\begin{align}
\label{Eq.:OffSiteContributions}
\text{Im}[S_{ss',j}(\bm{r})]=&\sum_{f\in \{A,B\}}4\zeta g_{j,f}^2\Bigl\{\cos(\varphi_{j,f})\nonumber\\
&+\sin[2\bm{q}_D\cdot(\bm{r}-\bm{R}_{j,f})]\Bigr\},\nonumber\\
\nonumber\\
\text{Re}[S_{ss',j}(\bm{r})]=&8 \zeta g_{j,A} g_{j,B}\Bigl\{\sin[\bm{q}_D\cdot(\bm{R}_{j,A}-\bm{R}_{j,B})]\nonumber\\
&\hspace{50pt}\times\sin[\frac{\varphi_{j,A}+\varphi_{j,B}}{2}]\nonumber\\
&+\cos[\bm{q}_D\cdot(2\bm{r}-\bm{R}_{j,A}-\bm{R}_{j,B})]\nonumber\\
&\hspace{50pt}\times\sin[\frac{\varphi_{j,A}-\varphi_{j,B}}{2}]\Bigr\}.
\end{align}
As for the on-site contribution, the label $ss'$ is kept only to indicate that it is an off-site contribution, while the vector $\bm{r}$ describes the position of the unit cell containing the sites $s$ and $s'$. The real part of the structure factor  is diagonal in the wave functions and contains only single MZM terms. The imaginary part is off-diagonal and proportional to the overlap between the wave functions of the two MZM-s that from the complex fermionic eigenmode. Inserting these results into Eq.~(\ref{Eq.:ZCorrelator}) and using that $S_{ss',j}^*=S_{s's,j}$, the dynamical structure factor reads 
\begin{align}
\label{Eq.:CorrelatorResult1}
S_{\bm{q}}(\omega)=&\frac{1}{\mathcal{N}}\sum_{j=1}^{N/2}\int d^2r\Bigl[S_{ss,j}(\bm{r})+\cos(a q_y)\text{Re}[S_{ss',j}(\bm{r})]\Bigr]\nonumber\\
&\times \Bigl[\delta(\omega-E_{\text{fl}}-E_{j})+\delta(\omega-E_{\text{fl}}+E_{k})\Bigr]\nonumber\\
&+\sin(a q_y)\text{Im}[S_{ss',j}(\bm{r})]\nonumber\\
&\times \Bigl[\delta(\omega-E_{\text{fl}}-E_{j})-\delta(\omega-E_{\text{fl}}+E_{k})\Bigr].
\end{align}

The formulas given by Eq.~(\ref{Eq.:OnSiteContribution}) and Eq.~(\ref{Eq.:OffSiteContributions}), together with Eq.~(\ref{Eq.:CorrelatorResult1}), provide, up to spatial integration, an analytic expression for the sub-gap dynamical structure factor $S_{\bm{q}}(\omega)$. In frequency space, it consists of isolated $\delta$-peaks distributed around the flux gap energy $\omega= E_{\text{fl}}$, because this is the energy that needs to be excited to switch the flux sectors from $N$ vortices to $N+2$ vortices. The exact position of these peaks in frequency space depends on the distribution of vortices and their distances to each other. In real systems however, the vortices will be distributed randomly. Therefore, we consider a quenched disorder average of the structure factor over the positions of the vortices in the next section.

\section{Structure factor in the presence of quenched disorder}
\label{Sec.:DisorderAverage}
We consider $N$ vortices which we divide into $N/2$ pairs ($\bm{R}_{j,A}, \bm{R}_{j,B}$), where $\bm{R}_{j,A}$ and $\bm{R}_{j,B}$ point to the vortex positions. The reason for this notation is explained below. We assume the positions of the vortices in the sample to be distributed randomly and thus apply a quenched disorder average over these positions. We treat these as random variables without dynamics caused by the Hamiltonian of the system. The average structure factor reads
\begin{align}
\bar{S}_{\bm{q}}(\omega)=&\frac{1}{\mathcal{N}}\int d^2R_{1,A}d^2R_{1,B}\cdots d^2R_{N/2,A}d^2R_{N/2,B}\nonumber\\
&\times P(\bm{R}_{1,A},\bm{R}_{1,B}\cdots\bm{R}_{N/2,A},\bm{R}_{N/2,B})S_{\bm{q}}(\omega),
\end{align} 
where $P(\bm{R}_{1,A}\cdots\bm{R}_{N/2,B})$ is the probability that a given vortex configuration is realized. 
Such a quenched disorder average smears out the $\delta$-peaks and introduces a natural broadening that we quantify in the following.

As the fermionic eigenmodes in the sub gap region depend on the distance between the vortices, we need to pair up the correct vortex bound states depending on the configuration $\bm{R}_{1,A},\bm{R}_{1,B}\cdots\bm{R}_{N/2,A},\bm{R}_{N/2,B}$. With the vortex density so low that at most two vortices are close to each other, we can approximately minimize the sum of the distances between all chosen pairs by pairing up a vortex at position $\bm{R}_{j,A}$ with its closest neighbor at $\bm{R}_{j,B}$ to form the eigenmodes $z_j$. This implies that the distribution $P(\bm{R}_{1,A}\cdots\bm{R}_{N/2,B})$ factorizes with $P(\bm{R}_{1,A}\cdots\bm{R}_{N/2,B})=P_2(\bm{R}_{1,A},\bm{R}_{1,B})\cdots P_2(\bm{R}_{N/2,A},\bm{R}_{N/2,B})$, where $P_2$ is a two-body probability distribution. It is defined by
\begin{align}
\label{Eq.:Distribution}
P_2(\bm{R}_A,\bm{R}_B)= \frac{1}{\pi^2\lambda^4 N} e^{-|\bm{R}_A-\bm{R}_B|^2/\lambda^2}=P_2(R),
\end{align}
and returns the probability of having a vortex at an arbitrary position $\bm{R}_{A}$ with its closest neighbor at distance $R=|\bm{R}_{A}-\bm{R}_{B}|$ from the point $\bm{R}_{A}$. It is normalized to $\int d^2R_Ad^2R_B P_2(\bm{R}_A,\bm{R}_B)=1$. The typical vortex spacing can be expressed by $\lambda=a\sqrt{L^2/\pi N}$ and is thus large for small vortex densities. We assume the latter to be so small that $\epsilon\lambda\gg1$, which assures that our assumption to take only two vortices close to each other into account is consistent.  We checked numerically that Eq.~(\ref{Eq.:Distribution}) corresponds indeed to the correct distribution of finding the closest point to another point in a 2D sample of randomly but uniformly distributed points. 

\begin{figure}
\includegraphics[scale=1.0]{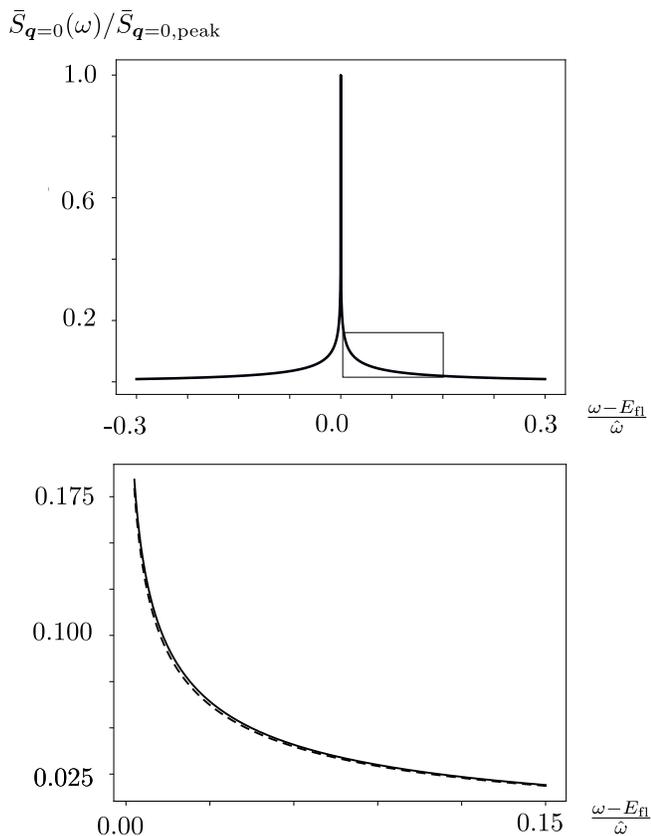}
\caption{\label{Fig.:DecayPlot} The upper panel shows the numerical results of Eq.~(\ref{Eq.:RemainingIntegralStructureFactor}) for the dynamical sub-gap structure factor normalized to the value at $\omega-E_{\text{fl}}=0$. The result is a narrow peak located around the flux gap $E_{\text{fl}}$. The lower panel shows a close up of the rectangle in the upper panel. The dashed line represents our analytic result $\bar{S}_{\bm{q}=0,\text{tail}}(\omega)$, while the solid line depicts the numerical result. We find an unconventional decay proportional to $\ln[\hat{\omega}/(\omega-E_{\text{fl}})\ln(\hat{\omega}/(\omega-E_{\text{fl}}))]^2$.}
\end{figure}

Each term in Eq.~(\ref{Eq.:CorrelatorResult1}) depends only on two vortex positions. Thus, we need to perform only two 2D non-trivial integrals. After averaging, each eigenmode labeled by $j$ in Eq.~(\ref{Eq.:CorrelatorResult1}) adds the same value to the final result, so that solving only one of the $N/2$ two-body integrals explicitly is sufficient. Therefore, we use $j=1$ for all expressions in the following. As the average distance between the vortices proportional to $\lambda$ is large compared to the lattice constant, we are interested in the structure factor at small $\bm{q}$. Due to the ultra short ranged correlators, there is only little structure in $\bm{q}$-space \cite{Knolle2015} and therefore we restrict our calculations on $\bm{q}=0$ for simplicity. Evaluated at $\bm{q}=0$, the structure factor can be measured in electron spin resonance. Its leading contribution is given by the on-site term $S_{ss,j}$ because $\text{Im}(S_{ss',j})$ is not relevant at $\bm{q}=0$ due the $\sin(aq_y)$ prefactor and $\text{Re}(S_{ss',j})$ is always sub dominant. The latter stems from the fact that the factor $g_{j,A}g_{j,B}$ is always exponentially smaller than one of the diagonal terms  $g_{j,A}^2$ or $g_{j,B}^2$. It remains to solve the integral
\begin{align}
\label{Eq.:FinalCorrelatorIntegral}
\bar{S}_{\bm{q}=0}(\omega)=&\frac{\zeta N}{2\mathcal{N}}\int d^2r d^2R_Bd^2R_AP_2(\bm{R}_{j,A},\bm{R}_{j,B})S_{ss,j}\nonumber\\
&\times \Bigl[\delta(\omega-E_{\text{fl}}-E_{k})+\delta(\omega-E_{\text{fl}}+E_{k})\Bigr].
\end{align}
As $\bm{r}$ appears only in differences with the vortex positions $\bm{R}_{j,A/B}$, the final result will be independent of the correlators position and thus we set $\bm{r}=0$ for each integral.
It is useful to introduce the relative vortex distance $\bm{R}=\bm{R}_{j,A}-\bm{R}_{j,B}$ and a center of mass coordinate $\bm{S}=(\bm{R}_{j,A}+\bm{R}_{j,B})/2$. The $\delta$-distributions as well as the probability density $P_2(R)$ are independent of the center of mass coordinate. Therefore, it is best to perform the center of mass integral first.
The second term in the on-site contribution of Eq.~(\ref{Eq.:OnSiteContribution}) proportional to $\sin[2\bm{q}_d\cdot(\bm{R}_{j,A/B})+\varphi_{j,A/B}]=\sin[2\bm{q}_d\cdot(\bm{S}\pm\bm{R}/2)+\varphi_{j,A/B}]$ oscillates fast on the scale $1/\epsilon$ for the center of mass integral while $g_{j,A/B}^2$ changes only slowly. Therefore, the oscillating term averages out and only the isotropic\footnote{In this case, isotropic means modulo the directional dependence of the energy splitting} first term in Eq.~(\ref{Eq.:OnSiteContribution}) remains. Note that if $\bm{R}_{j,A/B}\propto\bm{e}_y$ there are no oscillations that can average out the contribution. However, this is only relevant for a single direction in a 2D system and thus the contribution to the final result is still small. 
Inserting the wave functions, the remaining integral in relative and center of mass coordinates reads 
\begin{align}
\label{Eq.:FinalCorrelatorIntegralApprox}
\bar{S}_{\bm{q}=0}(\omega)=&(\zeta N\epsilon/4\pi)\int d^2rd^2S\;d^2R\;P_2(R)\nonumber\\
&\times \biggl(\frac{e^{-\epsilon |\bm{S}+\bm{R}/2|}}{|\bm{S}+\bm{R}/2|}+\frac{e^{-\epsilon |\bm{S}-\bm{R}/2|}}{|\bm{S}-\bm{R}/2|}\biggr)\nonumber\\
&\times\Bigl[\delta(\omega-E_{\text{fl}}-E_{k})+\delta(\omega-E_{\text{fl}}+E_{k})\Bigr]\nonumber\\
=&(\zeta N/2)\int d^2r d^2R P_2(R) e^{-\epsilon R}\nonumber\\
&\times\Bigl[\delta(\omega-E_{\text{fl}}-E_{k})+\delta(\omega-E_{\text{fl}}+E_{k})\Bigr].
\end{align}
In the second step, we performed the center of mass integral $dS^2$ which is independent of the $\delta$-distributions and the probability density $P_2(R)$. In the remaining integral, the $\delta$-distributions contribute for vortex separations $\bm{R}$ that satisfy $\omega-E_{\text{fl}}\pm \Delta(\bm{R})=0$.
From this, we find the angle integral $d\varphi_R$ from $d^2R=R dR d\varphi_R$ in the continuum limit to result in the factor
\begin{align}
\label{Eq.:IFact}
I(R,\omega)=
\begin{cases}
\frac{2}{\sqrt{[\Delta_P(R)]^2-(\omega-E_{\text{fl}})^2}}, \hspace{10pt}& |\omega-E_{\text{fl}}|<|\Delta_P(R)|,\\
0 ,\hspace{10pt}&\text{otherwise},
\end{cases}
\end{align}
where the radial symmetric $\Delta_P(R)=\Delta(\bm{R})/|\cos(\bm{q_D}\cdot\bm{R})|$ is the energy splitting between vortices without the direction depended oscillations, see Appendix \ref{App.:AngleInt}. The remaining integral 
\begin{align}
\label{Eq.:RemainingIntegralStructureFactor}
\bar{S}_{\bm{q}=0}(\omega)=&\frac{\zeta N}{2\mathcal{N}}\int d^2r\;\int_0^{R_C}  dR R P_2(R)I(R,\omega)e^{-\epsilon R},
\end{align}
can be solved numerically and leads to a peak in the dynamical structure factor at $\omega=E_{\text{fl}}$, see Fig.~\ref{Fig.:DecayPlot}. This peak is the signature of the zero modes inducing the $2^{N/2}$-fold degenerate ground state.

To provide analytical results describing the peak, we note that the factor $I(R,\omega)$ introduces a cutoff at
\begin{align}
R_C(\omega)=\frac{\ln\{\hat{\omega}/[\omega-E_{\text{fl}}]\ln[\hat{\omega}/(\omega-E_{\text{fl}})]\}}{2\epsilon},
\end{align}
which is the approximate solution of the equation $(\omega- E_{\text{fl}})^2=\Delta_P(R)^2$ valid for $\omega\ll\hat{\omega}$ with
\begin{align}
 \hat{\omega}=8v \epsilon/\pi^{1/2}.
\end{align}
The height of the peak can easily be found by evaluating Eq.~(\ref{Eq.:RemainingIntegralStructureFactor}) at $\omega- E_{\text{fl}}=0$.  We find
\begin{align}
\label{Eq.:TopPeak}
\bar{S}_{\bm{q}=0,\text{peak}}=\frac{v\zeta\Gamma(5/4)}{2(\pi \lambda)^{3/2}J^2\epsilon^{1/2}}\approx 0.08\frac{v\zeta}{J^2\lambda^{3/2}\epsilon^{1/2}},
\end{align}
with $\Gamma(x)$ being the Gamma function. To find an analytic expression for the shoulder of the peak and its decay behavior we use that for the decay the integrand is dominated at $R=R_C$. We expand the factor under the square root in $I(R,\omega)$ around $R_C$ up to first order and replace the remaining $R$ variables by the dominating $R_C$. Here, we find
\begin{align}
\label{Eq.:Tail}
\bar{S}_{\bm{q}=0,\text{tail}}(\omega)=&\frac{v \zeta \ln[\hat{\omega}/(\omega-E_{\text{fl}})\ln(\hat{\omega}/(\omega-E_{\text{fl}}))]^2}{2^{1/2}\pi^{3/2}J^2\epsilon^3\lambda^4},\nonumber\\
\approx&\frac{0.13 v \zeta \ln[\hat{\omega}/(\omega-E_{\text{fl}})\ln(\hat{\omega}/(\omega-E_{\text{fl}}))]^2}{J^2\epsilon^3\lambda^4}.
\end{align}
The results show that the shoulder of the peak has an unusual decay proportional to $\ln[\hat{\omega}/(\omega-E_{\text{fl}})\ln(\hat{\omega}/(\omega-E_{\text{fl}}))]^2$. This interesting feature and the position of the peak centered around $E_{\text{fl}}$ are signatures for the presence of MZM-s in the system. We compare the analytic results to the numerical results in Fig.~\ref{Fig.:DecayPlot}. 

The decay of the structure factor depends on the five important parameters $J,\epsilon, \lambda, v$ and $\zeta$. The material parameter $v$ depends on the lattice spacing and the energy per bond and is therefore constant for a given material. The parameter $\epsilon$ can be tuned by the magnetic field while $\lambda$ contains the information about the density of vortices present. The parameter $\zeta$ depends on the choice of $J_x, J_y$
and $J_z$. Our results are obtained for the isotropic point $J_x=J_y=J_z=J$ and therefore, we have a constant $\zeta\approx 0.8$.
\section{Conclusion}
\label{Sec.:Conclusion}
In the present work, we have analyzed the properties of the non-Abelian phase of Kitaev's honeycomb model in the presence of an external magnetic field. First, we derived the continuum Hamiltonian describing the system. Using this Hamiltonian, we analytically computed the wave functions of the MZM-s attached to vortex defects. These wave functions decay proportional to the inverse square root of the distance to the vortex position additionally to an exponential decay. The inverse decay length $\epsilon$ is determined by the ratio of the strength of the magnetic field and the spin-spin interaction. Furthermore, we obtained an analytic expression for the energy splitting that arises when two vortices approach each other which results in hybridization of the zero modes. We find that the energy splitting also decays as a function of distance between the vortices with a power law dependence, in addition to the exponential decay. On top of this decay, we found an oscillating prefactor $|\cos(\bm{q_D}\cdot\bm{R})|$ of the splitting that depends not only on the distance between the vortices but also on the direction. 
Using these results, we calculated a quenched disorder average of the low energy dynamical structure factor in presence of vortices. This quantity is relevant for experiments attempting to realize this model and has an unique signature of the presence of bound MZM-s. It adds a peak to the structure factor at the position of the flux gap $E_{\text{fl}}$ which is a direct result of the approximate ground state degeneracy due the MZM-s. Moreover, this peak shows an unconventional decay behavior proportional to $\ln[\hat{\omega}/(\omega-E_{\text{fl}})\ln(\hat{\omega}/(\omega-E_{\text{fl}}))]^2$ that could be used to characterize the Kitaev interactions in a material. As the existence of the additional peak at $E_{\text{fl}}$ only depends on the ground state degeneracy, it will be stable against perturbations to the Kitaev Hamiltonian as long as the system hosts MZM-s. However, it remains for future work to show if the characteristic decay of the peak we derived in this work persists in the presence of small perturbations to the Kitaev Hamiltonian. 

\section{Acknowledgments}
D.O. and F.H. acknowledge support from the Deutsche Forschungsgemeinschaft (DFG) under Grant No. HA 7084/2-1.
A.R. acknowledges the support of the Alexander von Humboldt foundation. The authors thank David DiVincenzo and Barbara Terhal for helpful discussions.
\newpage
\appendix
\section{Derivation of Angle Integral}
\label{App.:AngleInt}
In this appendix, we derive the expression $I(R,\omega)$ of Eq.~(\ref{Eq.:IFact}). We want to solve the angle integral
\begin{align}
I(R,\omega)=&\int d\varphi_R \;\delta(\omega-E_{\text{fl}}-E_{k}),
\end{align} 
where we only considered one of the $\delta$-distributions. This is possible, because one of the $\delta$-distributions only contributes for positive $\omega-E_{\text{fl}}$ while the other has the same contribution for negative $\omega-E_{\text{fl}}$. The angle $\varphi_R$ is the angle between $\bm{q}_D$ and $\bm{R}$.
We have to evaluate the $\delta$-distribution for the argument 
\begin{align}
\omega-E_{\text{fl}}- \Delta_P(R)|\cos[|\bm{q}_D|R\cos(\varphi_R)]|.
\end{align}
For large $|\bm{q}_D|R$, the $\cos$-factor oscillates $4 |\bm{q}_D|R/\pi$ times between $1$ and $-1$ for $\varphi_R \in [0,2\pi[$. We substitute $z_m=\Delta_P(R)|\cos[|\bm{q}_D|R\cos(\varphi_R)]$ on every monotonous part $m$ of the $\cos$-factor and find
\begin{align}
I(R,\omega)=&\sum_{m}\int^1_{-1} dz_m\;\frac{\delta(\omega-E_{\text{fl}}-z_m)}{|\bm{q}_D|R\sqrt{\Delta^2_P(R)-z_m^2}|\sin(\varphi_R)|}.
\end{align}
The factor $|\sin(\varphi_R)|$ is almost constant on a each part $m$ and for different $m$ all values are visited uniformly. Therefore, we can approximately replace $\sin(\varphi_R)$ by the average $2/\pi$ and obtain
\begin{align}
\label{Eq.:ResultIFac}
I(R,\omega)=&\sum_{m}\int dz_m\;\frac{\pi\delta(\omega-E_{\text{fl}}-z_m)}{2|\bm{q}_D|R\sqrt{\Delta^2_P(R)-z_m^2}}\nonumber\\
=& \frac{2}{\sqrt{\Delta^2_P(R)-(\omega-E_{\text{fl}})^2}},
\end{align}
In this step, we used the fact that every of the $4 |\bm{q}_D|R/\pi$ terms in the sum has the same contribution. 
Note that in this derivation, we ignored that there is a divergence if $\sin(\varphi_R)=0$, so that $\omega=\omega_d=\cos(|\bm{q}_D|R)$. This divergence leads to a peak in Eq.~(\ref{Eq.:ResultIFac}) at $\omega_d$. However, we find that the area under the peak is $\propto 1/\sqrt{|\bm{q}_D|R}$, which vanishes for large $|\bm{q}_D|R$.

\end{document}